\documentclass{raa_twocolumn}

\usepackage{graphicx,times}             
\usepackage{natbib}
\usepackage{amssymb,amsmath}

\bibpunct{(}{)}{;}{a}{}{,}

\usepackage[
colorlinks=true,        
citecolor=blue,         
linkcolor=blue,         
urlcolor=blue,          
pagebackref=true
]{hyperref}             

\begin{document}

  \title{Inflation in a nutshell: From basics to latest advances}
  
  \volnopage{Vol.0 (202x) No.0, 000--000}      
  \setcounter{page}{1}          
  
  \author{Zhi-Chong Qiu 
  	\inst{1,2}
  	\and Qing-Guo Huang
  	\inst{1,2,3}
  }
  
  \institute{Institute of Theoretical Physics, Chinese Academy of Sciences, Beijing 100190, China; {\it qiuzhichong@itp.ac.cn,huangqg@itp.ac.cn}\\
  	\and
  	School of Physical Sciences, 
  	University of Chinese Academy of Sciences, 
  	No. 19A Yuquan Road, Beijing 100049, China\\
  	\and
  	School of Fundamental Physics and Mathematical Sciences, Hangzhou Institute for Advanced Study, UCAS, Hangzhou 310024, China\\
  	\vs\no
  	{\small Received 202x month day; accepted 202x month day}}

\abstract{ Inflation is an elegant paradigm for the very early Universe. It not only offers a simple solution to the flatness and horizon puzzles of the standard hot Big Bang model, but also generates quantum fluctuations that seed CMB anisotropies and the formation of large-scale structure. In particular, both the spatial flatness of the Universe and a nearly Gaussian, scale-invariant power spectrum of the curvature perturbation predicted by inflation have been confirmed by various observations. Recently, a larger spectral index of curvature perturbation is preferred when combining with ACT DR6, and particularly when further including the $H_0$ prior from SH0ES, and then the Starobinsky inflation model is disfavored at more than $95\%$ confidence level. Even though modified reheating histories and non‑minimal couplings have been proposed to achieve a larger value of the spectral index, the model with higher‑order curvature corrections to Starobinsky inflation offers a concise and well‑motivated explanation.
}

   \authorrunning{Z.-C. Qiu \& Q. G. Huang }            
   \titlerunning{Inflation in a nutshell }  
   \maketitle
\section{Introduction}           
\label{sect:intro}

The standard hot Big Bang cosmology provides a successful framework for describing the thermal history of the Universe after the onset of the radiation-dominated era. In particular, it accounts for the cosmic microwave background (CMB) as relic radiation from the primordial plasma, last scattered when photons decoupled after recombination, and for the observed primordial abundances of light elements through Big Bang nucleosynthesis (BBN). Nevertheless, when extrapolated to very early times, this framework leaves several questions about the required initial conditions unresolved, most notably the flatness and horizon problems~\citep{Guth:1980zm}.

Inflation was proposed in the early 1980s~\citep{Starobinsky:1980te,Guth:1980zm,Linde:1981mu,Albrecht:1982wi} as an early stage of accelerated expansion capable of addressing several puzzles in the hot Big Bang. Beyond its role in background evolution, inflation provides a mechanism for generating primordial perturbations from quantum fluctuations~\citep{Mukhanov:1981xt,Hawking:1982cz,Starobinsky:1982ee,Guth:1982ec,Bardeen:1983qw}. In the scalar sector, canonical single-field slow-roll models typically predict predominantly adiabatic and nearly Gaussian curvature perturbations, with a nearly scale-invariant power spectrum~\citep{Sasaki:1986hm,Mukhanov:1988jd,Mukhanov:1990me,Maldacena:2002vr,Linde:2007fr,Baumann:2009ds}. These generic features are broadly consistent with current CMB observations~\citep{Planck:2018jri,Planck:2019kim}. Inflation can also generate a primordial tensor component~\citep{Starobinsky:1979ty} and a key observational target is the associated imprint on CMB $B$-mode polarization~\citep{Seljak:1996gy,Kamionkowski:1996ks}. No evidence for such a primordial tensor component has been found so far, and current CMB $B$-mode measurements place upper limits on its amplitude~\citep{BICEP:2021xfz}.

Over the past four decades, inflation has developed into a leading paradigm for the very early Universe. A large number of inflationary models have been proposed~\citep{Martin:2013tda,Martin:2013nzq}, and representative models provide useful reference points for clarifying the underlying physical mechanisms and comparing theoretical predictions with observations. In practice, observational constraints on $n_s$, inferred primarily from CMB anisotropy measurements, together with upper limits on the tensor-to-scalar ratio $r$ from the non-detection of primordial tensor modes, have progressively restricted the inflationary parameter space, disfavoring some simple convex monomial potentials with large $r$ while leaving several broad classes phenomenologically viable~\citep{Planck:2018jri,BICEP:2021xfz}.

While there are many excellent reviews of inflationary cosmology~\citep{Lyth:1998xn,Riotto:2002yw,Kachru:2003sx,Linde:2007fr,Baumann:2009ds,Mazumdar:2010sa,Wang:2013zva,Vazquez:2018qdg,Achucarro:2022qrl,Odintsov:2023weg,Ellis:2023wic,Kallosh:2025ijd}, this review aims to provide a focused overview of the physical logic connecting slow-roll dynamics, primordial perturbations, representative model predictions, and the latest observational constraints. We give a brief review of slow-roll inflation in Sec. \ref{Sec_II} and then discuss several representative models, including large-field, small-field, and plateau scenarios, in Sec.~\ref{Sec_III}. In Sec.~\ref{Sec_IV}, we discuss the impact of recent observations on inflation and future observational probes. Finally, we present our summary and discussion in Sec.~\ref{Sec_V}.


\section{A brief review of slow-roll inflation}
\label{Sec_II}

\subsection{Motivations for the Inflationary Paradigm}

The standard hot Big Bang cosmology provides a successful description of the thermal history of the Universe once it is in a hot, dense, and nearly homogeneous and isotropic state. However, it does not by itself explain why the early Universe should have satisfied such special initial conditions~\citep{Riotto:2002yw,Baumann:2009ds,Wang:2013zva}. In particular, during the radiation- and matter-dominated epochs, the dominant cosmic components satisfy $\rho+3p>0$, implying decelerated expansion. Consequently, the comoving Hubble radius $(aH)^{-1 }$ increases monotonically with time.

One issue is the flatness problem. Using the Friedmann equation, the total density parameter $\Omega_{\text{tot}}$ is related to the spatial curvature constant $K$ by
\begin{equation}
	\Omega_{\text{tot}}(t)-1={K \over (aH)^{2 }}\,,
\end{equation}
so during a decelerating expansion, any nonzero value of $|\Omega_{\rm tot}-1|$ grows as the comoving Hubble radius increases. The observed near-flatness of the Universe today, $|\Omega_{0}-1|\ll 1$~\citep{Planck:2018jri}, therefore implies that the early Universe must have been extraordinarily close to spatial flatness. A conventional extrapolation back to the Planck epoch gives the familiar order-of-magnitude estimate of order $|\Omega_{\rm tot}-1|\lesssim 10^{-60}$ in the absence of an earlier mechanism that drives the Universe toward flatness~\citep{Riotto:2002yw,Baumann:2009ds}.

A second issue is the horizon problem. In a decelerating universe, the comoving particle horizon $\tau=\int(aH)^{-1 }d\ln a$ is finite at any finite time. Near the epoch of recombination, at $z_{*}\simeq 1090$, the Thomson visibility function peaks and the CMB photons observed today last scattered from a finite-thickness last-scattering surface. The causal scale on the last-scattering surface subtends only an angle of order $1^\circ$ on the sky. Thus, without an earlier accelerated stage the observed last-scattering surface contains $\mathcal{O}(10^{4 })$ causally disconnected patches within the standard hot Big Bang evolution, while the CMB temperature anisotropies across the sky are only of order $\Delta T/T\sim \mathcal{O}(10^{-5})$~\citep{COBE:1992syq,Planck:2018vyg}. The standard hot Big Bang cosmology also leaves other initial condition and relic abundance questions unresolved, such as the monopole problem and the origin of the large entropy of the observable Universe~\citep{Riotto:2002yw,Wang:2013zva}. Moreover, it does not explain the origin of the primordial perturbations that later seed the CMB anisotropies and large-scale structure (LSS)~\citep{Mukhanov:1981xt,Bardeen:1983qw}.

These shortcomings can be addressed by a period of accelerated expansion before the standard radiation-dominated era, known as cosmic inflation. In a Friedmann-Robertson-Walker spacetime, accelerated expansion requires $\ddot{a}>0$ which is equivalent to an effective equation of state $w_{\rm eff}<-1/3$ for the total cosmic fluid. In the de Sitter limit, the cosmic expansion is dominated by an effective cosmological constant with equation of state $w_\Lambda\simeq -1$, so that
\begin{equation}
	H \simeq \text{const.}\,,\quad a(t) \simeq a_{i}{e}^{ H(t-t_{i}) }\,.
\end{equation}
During such a quasi-de Sitter phase, the comoving Hubble radius decreases approximately as
\begin{equation}
	(aH)^{-1}\propto e^{-\Delta N_i}\,,
\end{equation}
where $\Delta N_i(t)\equiv \ln\frac{a(t)}{a_i}$ denotes the number of $e$-folds elapsed since the beginning of inflation. The curvature contribution is exponentially diluted during quasi-de Sitter expansion as
\begin{equation}
	|\Omega_{\text{tot}}-1| \propto e^{-2\Delta N_i} \,,
\end{equation}
thus even an initially non-negligible curvature can be driven to an unobservably small value. In other words, the spatial flatness of our Universe -- one of the key predictions of inflation -- has been confirmed by observations~\citep{Planck:2018jri}.

Inflation also changes the causal structure of spacetime: a region that was initially in causal contact can be stretched to encompass the entire present observable Universe. A convenient order-of-magnitude estimate is that the present comoving Hubble scale was smaller than the comoving Hubble radius at the beginning of inflation
\begin{equation}
	(a_i H_i)^{-1}\gtrsim (a_0H_0)^{-1}\,.
\end{equation}
During quasi-de Sitter expansion, this condition may be written as
\begin{equation}
	(a_{\rm end}H_{i})^{-1}e^{ N_{\rm e}} \gtrsim (a_{0}H_{0})^{-1}\,,
\end{equation}
where $N_{\rm e} = \ln({a_{\rm end}}/{a_i})$, $H_i\simeq H_{\rm end}$ and the subscript end denotes the end of inflation. Assuming that the Universe subsequently enters the standard hot Big Bang thermal history at an energy scale $E_{\text{inf}}$, with approximate entropy conservation $aT\simeq \text{const.}$ and $H_i\sim E_{\text{inf}}^2/M_p$, one obtains the order-of-magnitude bound estimate
\begin{equation}
	N_{\rm e} \gtrsim \ln \left( \frac{T_0}{H_0}\frac{E_{\text{inf}}}{M_p} \right)\,.
\end{equation}
For $T_0\sim 10^{-4}\,{\rm eV}$, $H_0\sim 10^{-33}\,{\rm eV}$ and a typical grand-unification scale $E_{\rm inf}\sim 10^{15}\,{\rm GeV}$, this gives $N_{\rm e}\sim 60$. The precise lower bound depends on the inflationary scale and the post-inflationary thermal history. Thus the usual requirement is not a fixed total duration, but a minimum duration of order several tens of $e$-folds, and the commonly quoted $N_*\simeq 50\text{--}60$ refers to the number of $e$-folds between the horizon exit of observable CMB modes and the end of inflation.

\subsection{The Dynamics of Slow-Roll Inflation}

As discussed above, a sufficiently long period of accelerated expansion can naturally address the horizon and flatness problems. However, a quasi-de Sitter phase driven by a strictly constant vacuum energy by itself does not provide a graceful exit to the standard hot Big Bang evolution~\citep{Linde:1981mu,Albrecht:1982wi,Baumann:2009ds}. This motivates a dynamical realization of inflation in terms of a slowly evolving scalar field $\phi$, the inflaton, whose potential energy drives a quasi-de Sitter stage for a finite duration. A minimal realization is provided by a canonical scalar field with action
\begin{equation}
	S = \int d ^{4} x \sqrt{ -g } \left[ \frac{M_{p}^{2 }}{2}R - \frac{1}{2}g ^{\mu \nu }\partial_{\mu} \phi\partial_{\nu}\phi - V(\phi) \right] \,,
\end{equation}
where the reduced Planck mass is defined by $M_p =1/\sqrt{8\pi G}$. In a spatially homogeneous and isotropic Friedmann-Robertson-Walker background with a homogeneous inflaton configuration $\phi(t,\mathbf{x})=\phi(t)$, the scalar field is equivalent to a perfect fluid with energy density and pressure
\begin{equation}
	\rho_{\phi} = \frac{1}{2}\dot{\phi}^{2 } + V(\phi)\,,\quad p_{\phi} = \frac{1}{2}\dot{\phi}^{2 }-V(\phi)\,.
\end{equation}
Varying the action with respect to the metric yields the Friedmann equations
\begin{equation}
	\begin{aligned}
		H ^{2 } + \frac{K}{a ^{2 }} &= \frac{1}{3M_{p}^{2 }}\rho_{\phi}\,, \\
		\frac{\ddot{a}}{a} & = - \frac{1}{6M_{p}^{2 }}\left( \rho_{\phi} + 3p_{\phi} \right) \,,
	\end{aligned}
\end{equation}
while variation with respect to $\phi$ gives the equation of motion for the inflaton
\begin{equation}
	\ddot{\phi} + 3 H \dot{\phi} + V_{,\phi}(\phi) = 0\,.
	\label{eq_SR_phi}
\end{equation}
From the acceleration equation, the scalar field drives accelerated expansion when
\begin{equation}
	\rho_\phi+3p_\phi <0\,,
\end{equation}
or equivalently $\dot\phi^2<V(\phi)$. In the stronger limit $\dot\phi^2\ll V(\phi)$, the equation of state satisfies $w_\phi \equiv p_\phi/\rho_\phi \simeq -1$, and the expansion becomes quasi-de Sitter with the scalar field slowly rolling down its potential. For this behavior to persist for a sufficiently large number of $e$-folds, the potential must be sufficiently flat. In the slow-roll regime, the field acceleration in Eq.~\eqref{eq_SR_phi} is negligible, and the potential gradient is approximately balanced by Hubble damping $3H\dot\phi\simeq -V_{,\phi}$. Together with potential domination, this gives the slow-roll conditions
\begin{equation}
	\frac{1}{2}\dot{\phi}^{2 }\ll V(\phi)\,,\quad |\ddot{\phi}|\ll 3H |\dot{\phi}|\,,
\end{equation}
under which the background equations reduce to
\begin{equation}
	H\simeq \sqrt{\frac{V}{3M_{p}^{2 }}}\,,\quad \dot{\phi}\simeq -\frac{V_{,\phi}}{3H}\,.
\end{equation}

To quantify the flatness requirements on the potential, it is convenient to introduce the potential slow-roll parameters:
\begin{equation}
	\epsilon_{V} \equiv \frac{M_{p}^{2 }}{2}\left( \frac{V_{,\phi}}{V} \right) ^{2 } \,,\quad \eta_{V}\equiv M_{p}^{2 } \frac{V_{,\phi \phi}}{V}\,,
	\label{SR_para_V}
\end{equation}
where the subscript $V$ indicates that these parameters are defined in terms of the potential. The slow-roll approximation requires $\epsilon_V\ll 1$ and $|\eta_V|\ll 1$ so that the potential energy dominates the background evolution and the field acceleration remains subleading. These conditions provide a useful framework for model construction.

Alternatively, to assess the background dynamics independently of the exact form of the potential, one can define the Hubble slow-roll parameters. In this review we use the convention
\begin{equation}
	\epsilon_{H} = -\frac{d \ln H }{d N} = -\frac{\dot{H}}{H ^{2 }}\,,\quad \eta_{H} = \frac{d \ln \epsilon_{H}}{d N}=\frac{\dot{\epsilon}_{H}}{H\epsilon_{H}}\,,
	\label{SR_para_H}
\end{equation}
where $N=\ln a$ is used as the time variable and the subscript $H$ denotes definitions based on the Hubble parameter. Kinematically, accelerated expansion is strictly equivalent to $\epsilon_H < 1$ since $\ddot{a}/a = H^2(1 - \epsilon_H)$. For slow-roll inflation with a sufficiently long period, $\epsilon_H$ must remain much smaller than unity and evolve slowly, requiring $|\eta_H| \ll 1$. During the slow-roll regime, these kinematic parameters are related to the potential parameters at leading order by
\begin{equation}
	\epsilon_{H} \simeq \epsilon_{V}\,,\quad \eta_{H}\simeq 4\epsilon_{V} - 2\eta_{V}\,.
\end{equation}

An important property of slow-roll inflation is its attractor behavior~\citep{Liddle:1994dx}. From an intuitive perspective, the potential energy acts as an effective vacuum energy, whereas the inflaton kinetic energy of deviations from slow-roll is rapidly redshifted by the exponential expansion. To see this, it is convenient to rewrite the inflaton equation of motion as
\begin{equation}
	\frac{d^{2 }\phi}{dN^{2 }}+(3-\epsilon_{H})\frac{d\phi}{dN}=-\frac{V_{,\phi}}{H^{2 }} \,,
\end{equation}
where $\frac{d \phi }{d N}=\dot{\phi} / H$ and the friction term proportional to $\frac{d\phi}{dN}$ dominates the damping of velocity perturbations. When the force term varies slowly, the decaying component of the field velocity approximately scales as
\begin{equation}
	\delta\dot{\phi} \propto  a^{-3 }\,,
\end{equation}
so the background trajectory quickly converges to the slow-roll solution $3H\dot{\phi}\simeq -V_{,\phi}$. In phase space, nearby homogeneous background velocities are therefore driven toward a common late-time trajectory. 

A successful inflation model requires not only a sustained period of slow-roll dynamics but also a graceful exit mechanism. As the field approaches a steeper region or the minimum of its potential, the slow-roll conditions eventually break down. After inflation ends, if the potential has a minimum, the inflaton typically undergoes coherent oscillations around it. Since inflation strongly dilutes pre-existing matter and radiation, a successful inflationary scenario needs to be followed by a post-inflationary reheating stage, during which the energy stored in the inflaton condensate is converted into particles and thermalized~\citep{Kofman:1997yn,Bassett:2005xm}, thereby recovering the hot radiation-dominated stage of the standard hot Big Bang cosmology.

\subsection{Primordial Perturbations and Generic Predictions of Slow-Roll Inflation}

Having established the slow-roll background dynamics, the next question is how inflation generates the primordial perturbations that seed the CMB anisotropies and the LSS. The key point is that the quasi-de Sitter background stretches microscopic quantum fluctuations from sub-Hubble to super-Hubble scales.

\begin{figure}[!htbp]
	\centering
	\includegraphics[width=0.95 \linewidth]{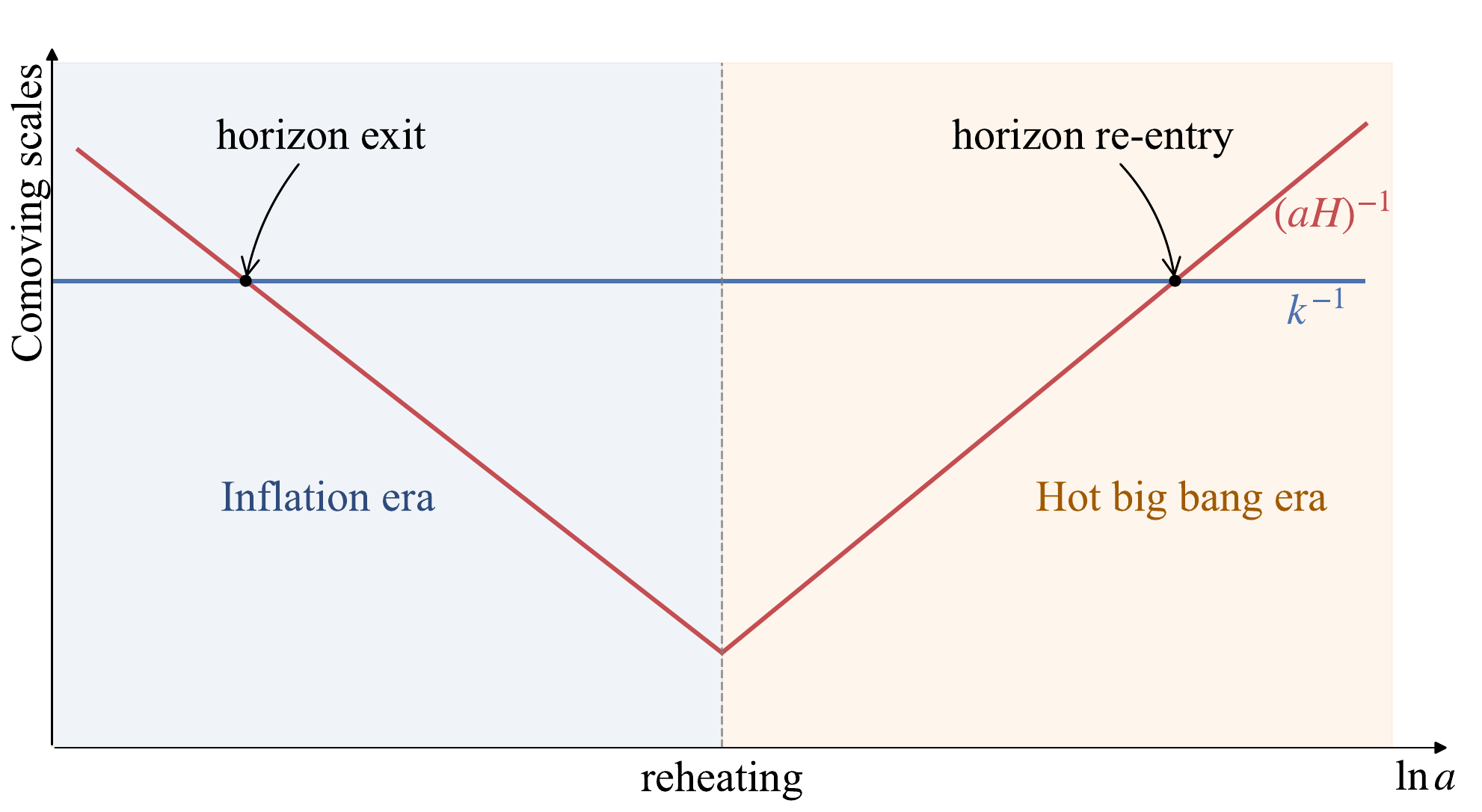}
	\caption{Comoving scales exit the horizon during inflation and re-enter during the hot Big Bang era.}
	\label{fig:inflation_horizon}
\end{figure}

For the canonical single-field inflation, after removing gauge redundancies and solving the constraint equations, the scalar sector contains only one propagating gauge-invariant degree of freedom, which may be taken to be the comoving curvature perturbation $\zeta$. Using the ADM formalism in unitary gauge~\citep{Maldacena:2002vr}, one obtains the quadratic action
\begin{equation}
	S^{(2) }_{\zeta} = \int dt d^{3 }x \, M_{p}^{2 }a^{3 }\epsilon_{H} \left[ \dot{\zeta}^{2 } - \frac{\left( \partial_{i}\zeta \right)^{2 } }{a^{2 }} \right].
\end{equation}
Introducing conformal time $\tau =\int dt/a$ and the Mukhanov-Sasaki variable~\citep{Sasaki:1986hm}
\begin{equation}
	u_{\mathbf{k}} = z_{s}\zeta_{\mathbf{k}}\,,\quad z_{s} = \sqrt{ 2 a^{2 }\epsilon_{H}M_{p}^{2 } }\,,
\end{equation}
the equation of motion in momentum space becomes
\begin{equation}
	u_{\mathbf{k}}^{\prime\prime } + \left( k^{2 }- \frac{z_{s}^{\prime\prime }}{z_{s}} \right)u_{\mathbf{k}} = 0\,, 
\end{equation}
where the prime denotes derivative with respect to $\tau$. On sub-Hubble scales, $|k\tau|\gg1$, the physical wavelength is much shorter than the Hubble radius, and the $k^{2 }$ term dominates the mode evolution. Choosing the Bunch-Davies vacuum therefore gives
\begin{equation}
	u_{\mathbf{k}} \simeq \frac{1}{\sqrt{ 2k }} e^{-ik\tau }\,,\quad \left| k\tau \right| \gg 1  \,.
\end{equation}
As the quasi-de Sitter expansion stretches the wavelength of the mode, the term $z_s''/z_s$ becomes dominant. In the slow-roll regime
\begin{equation}
	\frac{z_{s}''}{z_{s}} \simeq \frac{\nu_{s}^{2 }- 1 / 4}{\tau ^{2 }}\,,
\end{equation}
where $\nu_{s} \simeq \frac{3}{2}+\epsilon_{H}+\frac{1}{2}\eta_{H}$, the solution that matches the Bunch-Davies vacuum at initial times is approximately
\begin{equation}
	u_{\mathbf{k}} \simeq \frac{\sqrt{ \pi }}{2}e^{i(\nu_{s}+{1\over 2}){ \pi \over 2} }\sqrt{ -\tau }H_{\nu_{s}}^{(1) }(-k\tau)\,,
\end{equation}
where $H_{\nu}^{(1)}$ is the Hankel function of the first kind. On super-Hubble scales, the decaying mode rapidly becomes negligible and the comoving curvature perturbation freezes out as $\zeta_k$ approaches a constant in the slow-roll attractor regime. The modes later re-enter the horizon during the radiation- or matter-dominated era, as shown in Fig.~\ref{fig:inflation_horizon}. In this way, quantum fluctuations generated on sub-Hubble scales during inflation are stretched to cosmic scales and provide the primordial initial conditions for CMB anisotropies and the formation of large-scale structure.

At leading order, the frozen amplitude may be evaluated at Hubble crossing, yielding
\begin{equation}
	\mathcal{P}_{\zeta}(k)  = \frac{k^{3 }}{2 \pi ^{2 }}\left| \frac{u_{\mathbf{k}}}{z_{s}} \right| ^{2 } \simeq \frac{H_{*}^{2 }}{8\pi ^{2 }M_{p}^{2 }\epsilon_{H*}}\left( \frac{k}{k_{*}} \right)^{n_{s}-1 }  \,,
\end{equation}
where quantities with a subscript $*$ are evaluated when the pivot mode satisfies $k_*=aH$, and $k_*$ denotes a reference pivot scale. The scale dependence is parameterized by the tilt~\citep{Stewart:1993bc,Sasaki:1995aw}
\begin{equation}
	\begin{aligned}
		n_s -1 &\simeq  -2\epsilon_H - \eta_H\\
		& \simeq -6\epsilon_V+2\eta_V\,.
	\end{aligned}
	\label{n_s}
\end{equation}
Since the tilt itself is already first-order in the slow-roll expansion, the running of this tilt is further suppressed at second order
\begin{equation}
	\alpha_s \equiv \frac{dn_s}{d\ln k} \simeq -2 \epsilon_H \eta_H -\xi_H\,,
\end{equation}
where $\xi_H = \frac{d \eta_H}{d N}$ and all quantities are evaluated at horizon exit.

Inflation can also generate primordial tensor modes, corresponding to gravitational waves sourced by fluctuations of the transverse-traceless part of the metric. Their amplitude is model dependent and is conventionally described by the tensor power spectrum and the tensor-to-scalar ratio. The tensor power spectrum is
\begin{equation}
	\mathcal{P}_{t}(k)\equiv \sum_{\sigma = +,\times}  \frac{k^3}{2\pi^2}|h_k^\sigma|^2\simeq \frac{2H^{2 }}{\pi ^{2 }M_{p}^{2 }}\left( \frac{k}{k_{*}} \right)^{n_{T} }\,,
\end{equation}
where $h_k^\sigma$ denotes the Fourier mode of the tensor perturbation with polarization $\sigma$ and the tensor spectral index is $n_{T}\simeq -2\epsilon_{H}$. The tensor-to-scalar ratio is defined by
\begin{equation}
	r\equiv \frac{\mathcal{P}_t(k_*)}{\mathcal{P}_{\zeta}(k_*)}\simeq 16\epsilon_{H*}\,.
\end{equation}
In canonical single-field slow-roll inflation, expressing $r$ in terms of the potential slow-roll parameter $\epsilon_{V}$ directly relates the tensor amplitude to the evolution of the scalar field. Equivalently, the Lyth relation implies that an observable value of $r$ requires a sufficiently large field excursion~\citep{Lyth:1996im}. If one imposes $\Delta\phi\lesssim M_p$ and assumes approximately constant $r$ over the observed interval, it gives the estimate
\begin{equation}
	r \lesssim 2.2 \times 10^{-3} \left( \frac{60}{\Delta N} \right)^{2} \,,
	\label{lyth}
\end{equation}
where $\Delta N=\left|N(\phi_{e})-N(\phi_{i})  \right|$ corresponds to the observable CMB scales. More precisely, in Ref.~\citep{Huang:2015xda}, accounting for the evolution of $r$ yields a more stringent upper limit: 
\begin{equation}
	r\lesssim 2(1-n_s)^2\left[e^{(1-n_s)\Delta N/2}-1\right]^{-2},
\end{equation}
which reduces to Eq.~(\ref{lyth}) in the limit of $n_s\rightarrow 1$.

An important property of canonical single-field inflation is the adiabatic evolution of primordial perturbations~\citep{Wands:2000dp,Weinberg:2003sw}. On super-Hubble scales, the evolution of the gauge-invariant curvature perturbation $\zeta$ obeys
\begin{equation}
	\dot{\zeta}=-\frac{H}{\rho+p}\,\delta p_{\rm nad}+ \mathcal{O}\!\left(\frac{k^2}{a^2H^2}\right)\,,
\end{equation}
where $\delta p_{\text{nad}}$ vanishes on large scales for purely adiabatic perturbations, and $\zeta$ is conserved up to gradient corrections.

Canonical single-field slow-roll inflation also predicts approximately Gaussian primordial perturbations~\citep{Maldacena:2002vr,Lyth:2005fi}. A local form for the curvature perturbation is often parameterized as
\begin{equation}
	\zeta(\mathbf{x}) = \zeta_{g}(\mathbf{x}) + \frac{3}{5}f_{\text{NL}}^{\text{local} }\left[ \zeta_{g}^{2 }(\mathbf{x})-\langle \zeta^{2 }_{g}(\mathbf{x})\rangle \right] +\cdots\,,
\end{equation}
where $\zeta_{g}(\mathbf{x})$ represents the Gaussian part of the curvature perturbation, and $f_{\text{NL}}^{\text{local} }$ characterizes the amplitude of local non-Gaussianity. The leading non-Gaussian signal is described by the bispectrum, defined by
\begin{equation}
	\begin{aligned}
		\langle  \zeta_{\mathbf{k}_{1}} \zeta_{\mathbf{k}_{2}} \zeta_{\mathbf{k}_{3}}  \rangle = \left( 2 \pi \right)^{3 }\delta ^{(3) }\left( \mathbf{k}_{1}+\mathbf{k}_{2}+\mathbf{k}_{3} \right)  \\
		\times  B_{\zeta}(k_{1},k_{2},k_{3}) \,,
	\end{aligned}
\end{equation}
where the bispectrum $B_{\zeta}$ is given by
\begin{equation}
	\begin{aligned}
		B_{\zeta}(k_{1},k_{2},k_{3}) &= \frac{6}{5}f_{\text{NL}}^{\text{local} }\left[ P_{\zeta}(k_{1})P_{\zeta}(k_{2})\right.  \\
		&\quad\left.   + P_{\zeta}(k_{2})P_{\zeta}(k_{3})+P_{\zeta}(k_{3})P_{\zeta}(k_{1})\right] \,,
	\end{aligned}
\end{equation}
where $P_{\zeta}(k)=\frac{2 \pi ^2}{k^3}\mathcal{P}_{\zeta}(k)$ is the dimensionful power spectrum. In canonical single-field slow-roll inflation with the Bunch-Davies vacuum, local-type non-Gaussianity $f_{\text{NL}}^{\text{local} }$ is suppressed by slow-roll parameters, namely
\begin{equation}
	f_{\text{NL}}^{\text{local} } = \frac{5}{12}(1-n_{s})\,.
\end{equation}
This relation holds up to slow-roll suppressed corrections~\citep{Maldacena:2002vr,Creminelli:2004yq,Cheung:2007sv}. Therefore, a robust detection of local-type non-Gaussianity with an amplitude significantly larger than the single-field consistency-relation expectation would point beyond the standard single-clock attractor scenario. Possible realizations include additional light fields with entropy-to-curvature conversion \citep{Gordon:2000hv}, non-attractor phases in which $\zeta$ continues to evolve on super-Hubble scales~\citep{Namjoo:2012aa,Martin:2012pe}, or non-standard initial states~\citep{Holman:2007na}.

\section{Representative inflationary models}
\label{Sec_III}

\textit{Planck} observations are broadly consistent with the generic scalar-sector predictions of standard single-field slow-roll inflation: the scalar power spectrum index is nearly scale-invariant but slightly red-tilted, with $n_s = 0.9649 \pm 0.0042$, and no statistically significant evidence has been found for primordial features or for local/equilateral non-Gaussianity~\citep{Planck:2018jri,Planck:2019kim}. The initial perturbations are also consistent with adiabatic initial conditions, with no compelling evidence for isocurvature modes~\citep{Planck:2018jri}. Tensor perturbations remain undetected. Combining Planck with BICEP/Keck 2015 (BK15) data gives the upper bound $r_{0.002}<0.056$~\citep{Planck:2018jri}, while the later BK18 analysis improved to $r_{0.05}<0.036$~\citep{BICEP:2021xfz}.

\begin{figure}[!htbp]
	\centering
	\includegraphics[width=0.95 \linewidth]{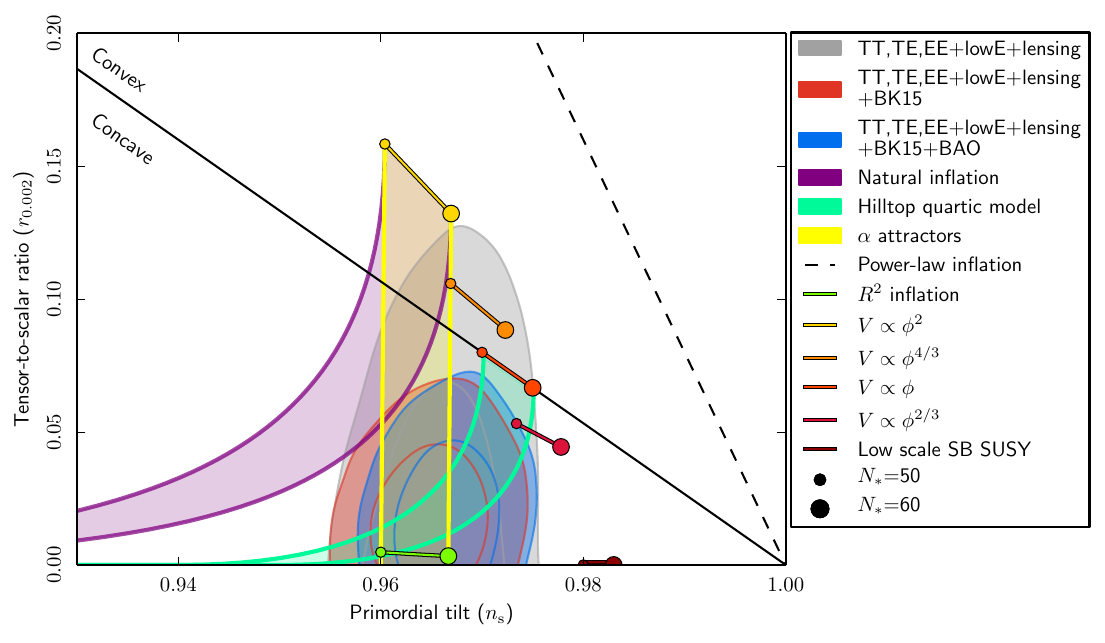}
	\caption{The figure reproduced from Ref.~\citep{Planck:2018jri} shows marginalized joint $68\%$ and $95\%$ CL regions in the $(n_s,r)$ plane, with $r$ evaluated at $k_{*}=0.002\;\mathrm{Mpc}^{-1}$. The inclusion of BK15 data substantially tightens the upper bound on $r$, thereby disfavoring simple large-field monomial potentials with large tensor amplitudes. Plateau-like and other concave potentials, including Starobinsky $R^{2}$ inflation, $\alpha$-attractors, and quartic hilltop models, remain compatible with these data for suitable choices of parameters and $N_*$.}
	\label{fig:planck2018}
\end{figure}

In this section, we introduce several typical inflationary models, grouped broadly into large-field, small-field, and plateau classes. Although \textit{Planck}-era CMB observations have strongly constrained several simple benchmark models, especially those with large tensor amplitudes, as shown in Fig.~\ref{fig:planck2018}, these models remain useful for illustrating the connection between the shape of the inflaton potential, slow-roll dynamics, and observable predictions. Modern inflationary model building often extends these benchmark scenarios in order to address theoretical issues or to accommodate increasingly precise observational constraints.

\subsection{Large-Field Inflation: Monomial and Natural Inflation}

For large-field inflation models, the inflaton typically undergoes a super-Planckian field excursion, $\Delta\phi \gtrsim M_p$ during the last $50 \text{--} 60$ $e$-folds of inflation. A representative example is the chaotic inflation scenario proposed by Linde~\citep{Linde:1983gd}. In this framework, inflation can start from a broad range of initial field values and kinetic energies, without requiring an initial state of thermal equilibrium. A simple class of chaotic models is described by the monomial potential
\begin{equation}
	V(\phi) = \lambda M_{p}^{4 } \left( \frac{\phi}{M_{p}} \right) ^{p }\,,
\end{equation}
where $\lambda$ is dimensionless in this normalization. The corresponding slow-roll parameters are given by
\begin{equation}
	\epsilon_{V} =\frac{p^2 M_p^2}{2\phi^2}\,, \quad \eta_V = \frac{p(p-1)M_p^2}{\phi^2}\,,
\end{equation}
where the field value at the end of inflation is $\phi_{\rm end} = p M_p / \sqrt{2}$, determined by $\epsilon_V(\phi_{\rm end})=1$. For $N_* \gg p / 4$, the number of $e$-folds $N_*$ from horizon exit $\phi_*$ to the end of inflation is
\begin{equation}
	N_* \simeq \frac{1}{M_p^2} \int_{\phi_{\rm end}}^{\phi_*} \frac{V}{V_{,\phi}} d\phi \simeq \frac{1}{2pM_p^2} \phi_*^2 \,,
\end{equation}
Substituting this relation into the slow-roll parameters yields
\begin{equation}
	n_{s}\simeq 1- \frac{p+2}{2N_{ *}}\,,\quad r \simeq \frac{4p}{N_{*}}\,.
\end{equation}

A closely related large-field scenario is natural inflation~\citep{Freese:1990rb,Adams:1992bn}, in which the inflaton is interpreted as a pseudo-Nambu--Goldstone boson (pNGB) with a periodic potential 
\begin{equation}
	V(\phi)=\Lambda^{4}\left[1+\cos\left(\frac{\phi}{f}\right)\right]\,,
\end{equation}
where $f$ is the decay constant and $\Lambda$ is an energy scale. The flatness of the potential is protected by an approximate shift symmetry. By defining
\begin{equation}
    s^{2 }_* \equiv \sin ^{2 }\left( \frac{\phi_*}{2f} \right) = \frac{2f^{2 }}{2f^{2 } + M_{p}^{2 }}\exp \left( -\frac{N_* M_{p}^{2 }}{f^{2 }} \right)\,,  
\end{equation}
the slow-roll parameters can be written as
\begin{equation}
    \epsilon_{V} = \frac{M_{p}^{2 }}{2f^{2 }} \frac{s_*^{2 }}{1-s_*^{2 }}\,,\quad \eta_{V}=- \frac{M_{p}^{2 }}{2f^{2 }} \frac{1-2s_*^{2 }}{1-s_*^{2 }}\,,
\end{equation}
and the scalar spectral index and tensor-to-scalar ratio are therefore
\begin{equation}
    n_{s}\simeq 1-\frac{M_{p}^{2 }}{f^{2 }} \frac{{1+s_{*}^{2 }}}{1-s_{*}^{2 }}\,,\quad r \simeq \frac{8M_{p}^{2 }}{f^{2 }} \frac{s_{*}^{2 }}{1-s_{*}^{2 }}\,.
\end{equation}
In the large-field limit $f \gg M_p$, natural inflation recovers the predictions of quadratic chaotic inflation as
\begin{equation}
	n_s \simeq 1 - \frac{2}{N_{*}}\,, \quad r \simeq \frac{8}{N_*}\,.
\end{equation}

Although these large-field models are conceptually simple and historically important, Fig.~\ref{fig:planck2018} shows that they are disfavored by the \textit{Planck} data, and more strongly when combined with BK15 data.

\subsection{Small-Field Inflation: Hilltop Models and D-Brane Inflation}

In contrast to large-field scenarios, small-field inflation models are characterized by a sub-Planckian field excursion, $\Delta \phi < M_p$. A representative paradigm of this class is the hilltop inflation~\citep{Boubekeur:2005zm}, which naturally arises in settings with spontaneous symmetry breaking. In this scenario, the inflaton rolls away from an unstable symmetric point at $\phi = 0$ toward a true vacuum. Near the local maximum, the effective potential is often parameterized as
\begin{equation}
	V(\phi) \simeq V_0 \left[ 1 - x^p \right] + \dots\,,
	\label{small_field_potential}
\end{equation}
where $x\equiv \phi / \mu$, $V_0$ is the vacuum energy density driving accelerated expansion, $\mu$ characterizes the width of the hilltop region and the ellipsis represents higher-order terms that stabilize the potential away from the hilltop region. For hilltop models, the leading power typically satisfies $p\geq 2$.

Another well-studied small-field realization is D-brane inflation motivated by string theory~\citep{Dvali:1998pa,Dvali:2001fw,Burgess:2001fx,Kachru:2003sx}. In the classic D3-$\overline{\text{D}3}$ brane scenario, the inflaton $\phi$ is related to the physical separation between a brane and an anti-brane moving toward each other in the extra dimensions. The attractive Coulomb-like interaction between the branes generates an effective potential of the inverse-power-law form which can be written schematically in the form of Eq.~\eqref{small_field_potential} with $p<0$. In this case, the field rolls from a locally flat region $x\gg 1$ toward smaller separation $\phi \to 0$, and inflation ends via tachyon condensation or brane annihilation~\citep{Sen:1998sm}.

At leading order in the flat-region expansion, $x^p\ll1$, the slow-roll parameters following from Eq.~\eqref{small_field_potential} are
\begin{equation}
	\begin{aligned}
		\epsilon_V &\simeq \frac{p^2 M_p^2}{2\mu^2} x^{2p-2}\,,\\
		\eta_V &\simeq -\frac{p(p-1)M_p^2}{\mu^2}x^{p-2}\,.
	\end{aligned}
\end{equation}
The quadratic hilltop case $p=2$ is special because $\eta_V$ is independent of $x$. The number of $e$-folds is given by
\begin{equation}
	N_* \simeq \frac{\mu^2}{2M_p^2}\ln\frac{x_{\rm end}}{x_*}\,,
\end{equation}
and hence
\begin{equation}
	\begin{aligned}
		n_{s} & \simeq 1- \frac{4M_{p}^{2 }}{\mu ^{2 }} \,,\\
		r & \simeq 32\frac{M_p^2}{\mu^2}x_{\rm end}^2 e^{-4M_p^2N_*/\mu^2} \,.
	\end{aligned}
\end{equation}
For both $p>2$ hilltop models and $p<0$ D-brane inflation, the number of $e$-folds is approximately
\begin{equation}
	N_{*} \simeq \frac{\mu ^{2 }}{p(p-2)M_{p}^{2 }} x_{*}^{2-p }\,,
	\label{small-field-N}
\end{equation}
where the contribution from $x_{\rm end}$ has been neglected. This gives
\begin{equation}
	\begin{aligned}
		n_{s} &\simeq 1 - \frac{2(p-1)}{(p-2)N_{*}}\,, \\
		r & \simeq 8 p ^{2 } \frac{M_{p}^{2 }}{\mu ^{2 }}\left[  \frac{p(p-2)M_{p}^{2 }}{\mu ^{2 }} N_{*} \right]^{-2 - {2 \over p-2} } \,.
	\end{aligned}
\end{equation}
These expressions show that $r$ is strongly suppressed in the flat-region limit: exponentially in the quadratic hilltop case and as a power law in the $p>2$ hilltop and $p<0$ brane cases. Therefore, these small-field models are less directly constrained by the current upper bound on $r$ alone and are broadly compatible with the \textit{Planck}+BK15 regions for suitable choices of $N_*$, as shown in Fig.~\ref{fig:planck2018}. However, this feature is accompanied by greater model dependence. The initial position and velocity of the inflaton must often be arranged so that the field does not overshoot the flat region.

\subsection{Plateau Models and Attractor Behavior}

Plateau potentials provide an important class of inflationary models in which the tensor amplitude is naturally suppressed while the scalar spectral index is predicted in a relatively robust way. Two representative examples are modifications of the gravitational sector, as in Starobinsky inflation~\citep{Starobinsky:1980te}, and the introduction of non-minimal couplings, as in Higgs inflation~\citep{Bezrukov:2007ep}. Despite their distinct physical origins, these models share an attractor behavior in their inflationary predictions~\citep{Kallosh:2013hoa}.

The action of a general $f(R)$ theory in the Jordan frame is
\begin{equation}
	S_{J} = \frac{M_{p}^{2}}{2} \int d^{4}x \sqrt{-g} f(R)\,.
\end{equation}
The Starobinsky inflation model corresponds to
\begin{equation}
	f(R)=R+\frac{R^{2 }}{2\mu^{2 }}\,,
\end{equation}
where $\mu$ is a mass scale. Introducing a non-dynamical auxiliary scalar field $\chi_s$, the action can be rewritten in a dynamically equivalent form
\begin{equation}
	S_{J} = \frac{M_{p}^{2}}{2} \int d^{4}x \sqrt{-g} \left[ f(\chi_s) + f^{\prime}(\chi_s)(R - \chi_s) \right]\,,
\end{equation}
with $\chi_s = R$ recovering the original $f(R)$ action for $f''(\chi_s)\ne 0$. Defining the conformal factor $\Omega_s^{2} \equiv f^{\prime}(\chi_s) = 1 + \frac{\chi_s}{\mu^{2}}$ and performing the conformal transformation $\tilde{g}_{\mu \nu} = \Omega_s ^{2 }g_{\mu \nu}$, the action in the Einstein frame takes the form
\begin{equation}
	S_{E} = \int d^{4}x \sqrt{-\tilde{g}} \left[ \frac{M_{p}^{2}}{2}\tilde{R} - \frac{1}{2}\tilde{g}^{\mu\nu}\partial_{\mu}\phi\partial_{\nu}\phi - V(\phi) \right]\,,
\end{equation}
where $\phi = \sqrt{ 6 }M_{p}\ln \Omega_s$, and the Starobinsky potential is
\begin{equation}
	\begin{aligned}
		V(\phi) &= \frac{M_{p}^{2}}{2} \frac{\chi_s f^{\prime}(\chi_s) - f(\chi_s)}{[f^{\prime}(\chi_s)]^{2}} \\
		&=\frac{1}{4} M_{p}^{2} \mu^{2} \left( 1 - e^{-\sqrt{2/3}\phi/M_{p}} \right)^{2}\,.
	\end{aligned}
\end{equation}
Thus, the $R^2$ correction in the Jordan frame can be reinterpreted in the Einstein frame as an additional scalar degree of freedom, whose potential approaches a flat plateau exponentially at large field values.

Higgs inflation realizes a plateau potential through a non-minimal coupling between the Higgs field and gravity~\citep{Bezrukov:2007ep}. In the unitary gauge and neglecting gauge interactions, the relevant Jordan-frame action is
\begin{equation}
	\begin{aligned}
		S_{J} = \int d^{4}x \sqrt{-g} \left[ \frac{M_{p}^{2} + \xi h^{2}}{2} R- \frac{1}{2}g^{\mu\nu}\partial_{\mu} h \partial_{\nu} h \right.\\
		\left.- \frac{\lambda}{4}(h^{2} - v^{2})^{2} \right]\,,
	\end{aligned}
\end{equation}
where $h$ denotes the Higgs field, $\xi$ is the non-minimal coupling and $v$ is the Higgs vacuum expectation value. During inflation $h\gg v$, the potential satisfies $V_{J}(h) \simeq \lambda h^{4} / 4$.  After the conformal transformation with $\Omega_h ^{2 }=1+\xi h ^{2 } / M_{p}^{2 }$, the gravitational sector is brought to the Einstein-Hilbert form at the price of generating a non-canonical kinetic term for $h$ in the Einstein frame. For a large non-minimal coupling $\xi \gg 1$ and in the inflationary regime $h \gg M_{p}/\sqrt{\xi}$, one introduces a canonically normalized field $\chi_h$ through
\begin{equation}
	\frac{d\chi_h}{dh} = \left[ \frac{1+\xi(1+6\xi)h^2/M_p^2}{(1+\xi h^2/M_p^2)^2} \right]^{1/2} \simeq \frac{\sqrt{6}M_p}{h}\,.
\end{equation}
The action in the Einstein frame is
\begin{equation}
	S_{E} = \int d^{4 }x\sqrt{ -\tilde{g} } \left[  \frac{M_{p}^{2 }}{2}\tilde{R} - \frac{1}{2} \tilde{g}^{\mu\nu}\partial_\mu\chi_h\partial_\nu\chi_h- V_{E}(\chi_h) \right] \,,
\end{equation}
and the Einstein frame potential of Higgs inflation has the same leading plateau form as the Starobinsky potential:
\begin{equation}
	V_{E}(\chi_h) \simeq \frac{\lambda M_{p}^{4}}{4\xi^{2}} \left( 1 - e^{-\sqrt{2/3}\chi_h/M_{p}} \right)^{2}\,,
\end{equation}
yielding the same leading-order large-$N_{*}$ predictions for $n_s$ and $r$.

The agreement between these apparently different models is not accidental, but rather a manifestation of a broader attractor behavior. In the Einstein frame, when $\varphi\gg M_p$, both Starobinsky and Higgs inflation approach a constant potential exponentially as~\citep{Kallosh:2013hoa,Ellis:2013nxa,Kallosh:2013tua,Kallosh:2013yoa}
\begin{equation}
	V(\varphi)=V_0\left(1-Ae^{-b\varphi/M_p}+\cdots\right)\,,
\end{equation}
where $A$ and $b$ are positive constants. For such a potential, the slow-roll parameters in the large-field limit give
\begin{equation}
	\epsilon_V\simeq {A^2b^2 \over 2} e^{-2b\varphi_*/M_p}\,, \quad \eta_V\simeq  - A b^2 e^{-b\varphi_*/M_p}\,,
\end{equation}
which implies $\epsilon_{V} \propto \eta_{V}^{2} \ll |\eta_{V}|$. Since the e-folding number before the end of inflation is evaluated as
\begin{equation}
	N_{\rm *} 
	\simeq \frac{1}{AbM_p}\int_{\varphi_{\rm end}}^{\varphi_*} e^{b\varphi/M_p}\,d\varphi
	\simeq \frac{1}{Ab^2} e^{b\varphi_*/M_p}\,,
\end{equation}
one obtains the universal large-$N_*$ behavior
\begin{equation}
	n_s\simeq 1-\frac{2}{N_*}\,,\quad r\simeq \frac{8}{b^2 N_{*}^2}\,,
	\label{plateau_ns_r}
\end{equation}
where $b = \sqrt{ 2 /3}$ for Starobinsky and Higgs inflation. For typical $N_{*} \in [50, 60]$, Starobinsky and Higgs inflation predict $n_{s} \simeq 0.960\text{--}0.967$ and a suppressed tensor amplitude $r \simeq 0.003\text{--}0.005$, in good agreement with the \textit{Planck}-era constraints as shown in Fig.~\ref{fig:planck2018}.

\section{Recent developments} 
\label{Sec_IV}
\subsection{Recent Observational Constraints on Inflation}

\begin{figure}[!htbp]
	\centering
	\includegraphics[width=0.95 \linewidth]{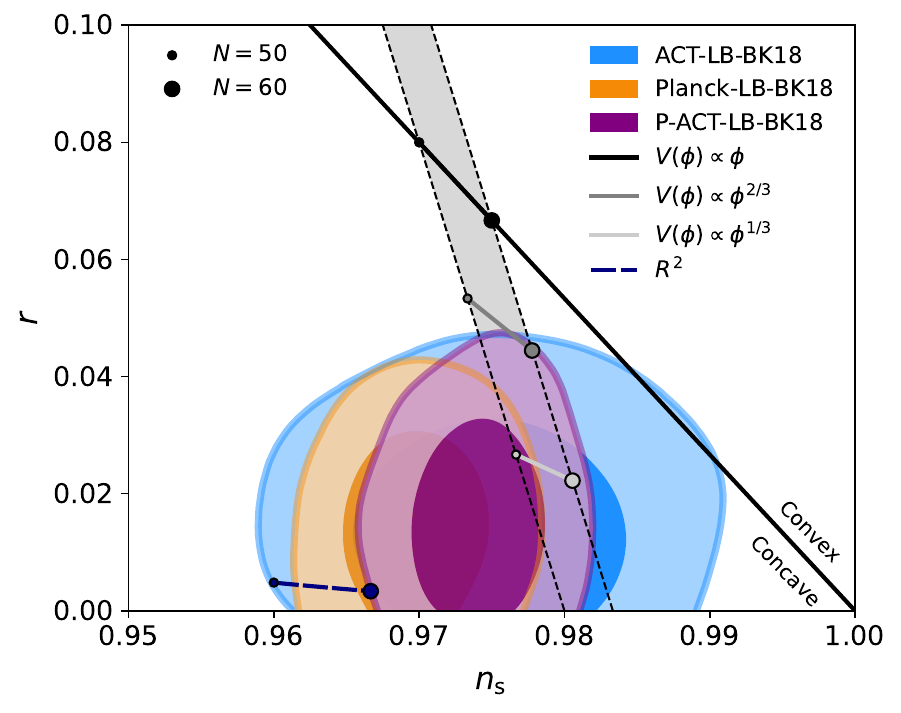}
	\caption{The ACT collaboration constraints on $n_{s}$ and $r$ at $k_*=0.05\;\text{Mpc}^{-1 }$ from the P-ACT-LB$_{\text{DESI DR1}}$-BK18 combination within the $\Lambda$CDM model~\citep{AtacamaCosmologyTelescope:2025nti}.}
	\label{fig:ACT}
\end{figure}

The latest ACT DR6 analyses~\citep{AtacamaCosmologyTelescope:2025blo,AtacamaCosmologyTelescope:2025nti}, especially when combined with selected CMB and BAO data, have been reported to prefer a larger value of $n_s$ compared with the \textit{Planck}-only preference. Within the $\Lambda$CDM model, the combined \textit{Planck}+ACT analysis gives $n_s = 0.9709 \pm 0.0038$, while including ACT and \textit{Planck} lensing, DESI DR1 BAO, and BK18 data further gives the P-ACT-LB$_{\text{DESI DR1}}$-BK18 constraint $n_s \simeq 0.9743 \pm 0.0034$, and a tight upper bound on the tensor-to-scalar ratio $r<0.038$~\citep{AtacamaCosmologyTelescope:2025nti,DESI:2024mwx,BICEP:2021xfz}. The SPT-3G D1 analysis also reported a larger value, $n_s=0.9728\pm0.0027$, for the \textit{Planck}, ACT, SPT (CMB-SPA)$+$DESI DR2 combination within the $\Lambda$CDM model~\citep{SPT-3G:2025bzu,DESI:2025zgx}. In these data combinations, conventional Starobinsky-like plateau models, including Starobinsky $R^2$ inflation and Higgs inflation in their minimal realizations, are placed under nontrivial pressure, as shown in Fig.~\ref{fig:ACT}. 

Several recent analyses have emphasized that the interpretation of this upward shift should be treated with caution~\citep{Ferreira:2025lrd,McDonough:2025lzo,Balkenhol:2025wms}. A recent combined analysis using CMB-SPA$+$BK18 data found $n_s=0.9682\pm0.0032$, while adding DESI DR2 BAO shifted the constraint to $n_s=0.9728\pm0.0029$~\citep{Balkenhol:2025wms}. As argued in Ref.~\citep{Ferreira:2025lrd}, since BAO data do not directly constrain $n_s$, this upward shift may arise from correlations between the CMB-inferred scalar tilt and BAO-related background parameters, particularly $r_d h$ and $\Omega_m$, together with the possible BAO--CMB tension within the $\Lambda$CDM model~\citep{Ferreira:2025lrd}. In addition, Ref.~\citep{McDonough:2025lzo} has also stressed the correlated shifts involving the optical depth $\tau_{\rm reio}$ and the role of CMB lensing and large-scale polarization choices. Therefore, whether the current data require a larger $n_{s}$ remains an open question.

\subsection{Recent Theoretical Responses to Data Combinations Preferring Larger $n_s$}

Although the observational status is not yet settled, the possibility that ACT DR6 and related CMB/BAO data combinations may prefer a larger scalar spectral index has already motivated renewed interest in inflationary model building. One possibility is to revisit the assumptions about the post-inflationary reheating history~\citep{Liddle:2003as,Dai:2014jja}. In fact, reheating affects the relation between the CMB pivot scale and the number of $e$-folds $N_*$ before the end of inflation, thereby shifting the predicted value of $n_s$~\citep{Drees:2025ngb,Zharov:2025zjg,Haque:2025uri,Liu:2025qca,Haque:2025uis,Maity:2025czp,Mohammadi:2025gbu,Odintsov:2025bmp}. The comoving pivot scale $k_{*}$ can be related to the present scale by
\begin{equation}
	\frac{k_{*}}{a_{0}H_{0}} = \frac{a_{*}}{a_{\rm end}} \frac{a_{\rm end}}{a_{\rm reh}} \frac{a_{\rm reh}}{a_{0}}\frac{H_*}{H_0} = e^{-N_{*} }e^{-N_{\rm reh} } \frac{a_{\rm reh}}{a_{0}}\frac{H_*}{H_0}\,,
\end{equation}
where $a_{\rm reh}$ denotes the scale factor at the end of reheating and $N_{\rm reh}\equiv \ln(a_{\rm reh}/a_{\rm end})$. Assuming a constant effective equation of state $w_{\rm reh}$ during reheating, the energy density during this epoch evolves as
\begin{equation}
	\rho\propto a^{-3(1+w_{\rm reh}) }\,.
\end{equation}
Together with entropy conservation after reheating and the relation
\begin{equation}
	\rho_{\rm reh} = \frac{\pi ^{2 }}{30}g_{\rm reh}T_{\rm reh}^{4 }\,,
\end{equation}
where $g_{\rm reh}$ is the effective number of relativistic degrees of freedom at reheating, the dependence of $N_{*}$ on the reheating history can be expressed as
\begin{equation}
	\begin{aligned}
		N_*  \simeq  61.6 -\ln\left(\frac{k_*}{0.05\,{\rm Mpc}^{-1}}\right) +\frac14  \ln\left(  \frac{V_*^2}{M_p^4\rho_{\rm end}}  \right)   \\
		+  \frac{1-3w_{\rm reh}}{12(1+w_{\rm reh})}  \ln\left(  \frac{\rho_{\rm reh}}{\rho_{\rm end}}  \right)\,,
	\end{aligned}
\end{equation}
where $V_*$ is the potential energy at horizon exit and $\rho_{\rm end}$ is the total energy density at the end of inflation. Since $\rho_{\rm reh}<\rho_{\rm end}$, the last term increases $N_*$ when $w_{\rm reh}>1/3$. Therefore, if the post-inflationary expansion is characterized by a stiff effective equation of state $w_{\rm reh}>1/3$, the predicted $n_{s}$ can shift to a larger value for plateau models.

A second response is to modify the gravitational sector. One route is to relax the attractor or strong-coupling limit in non-minimally coupled scalar-tensor models~\citep{Kallosh:2025rni} in which the quadratic chaotic model with a linear non-minimal coupling to gravity:
\begin{equation}
	\begin{aligned}
		S = \int d^{4 }x \sqrt{ -g }\left[ \frac{M_{p}^{2 }}{2}\left( 1+\frac{\xi \phi}{M_{p}} \right)R-\frac{1}{2}g^{\mu\nu}\partial_{\mu}\phi \partial_{\nu}\phi \right. \\
		\left. -\frac{1}{2}m ^{2 }\phi ^{2 } \right] \,.
	\end{aligned}
\end{equation}
For $\xi = \mathcal{O}(1)$, its large-$N_*$ predictions approach
\begin{equation}
	n_{s}\simeq 1- \frac{3}{2N_*}\,,\quad r \simeq \frac{4}{\xi N_*^{ 3 /2}}\,,
\end{equation}
which have been reported to be compatible with the larger $n_s$ P-ACT-LB constraints for $N_*=60$ and $0.3\lesssim \xi \lesssim 4$~\citep{Kallosh:2025rni}.

\begin{figure}[!htbp]
	\centering
	\includegraphics[width=0.95 \linewidth]{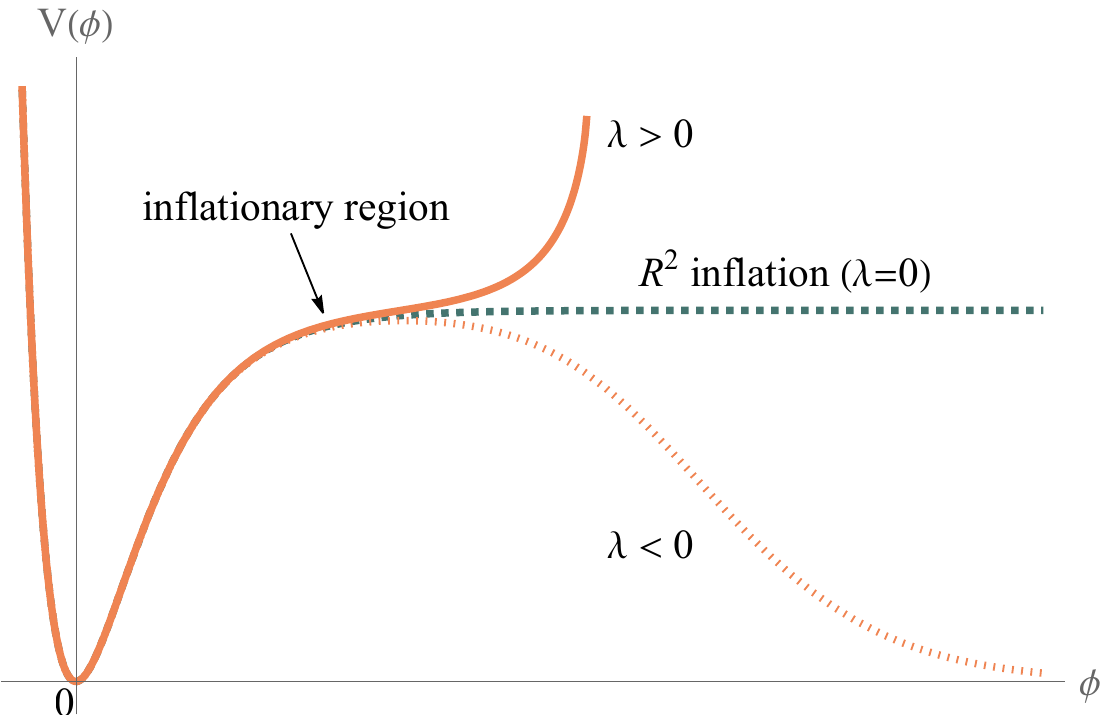}
	\caption{Schematic Einstein frame effective potential of the non-perturbative exponential $f(R)$ model for different values of $\lambda$.}
	\label{fig:fR_potential}
\end{figure}

\begin{figure}[!htbp]
	\centering
	\includegraphics[width=0.95 \linewidth]{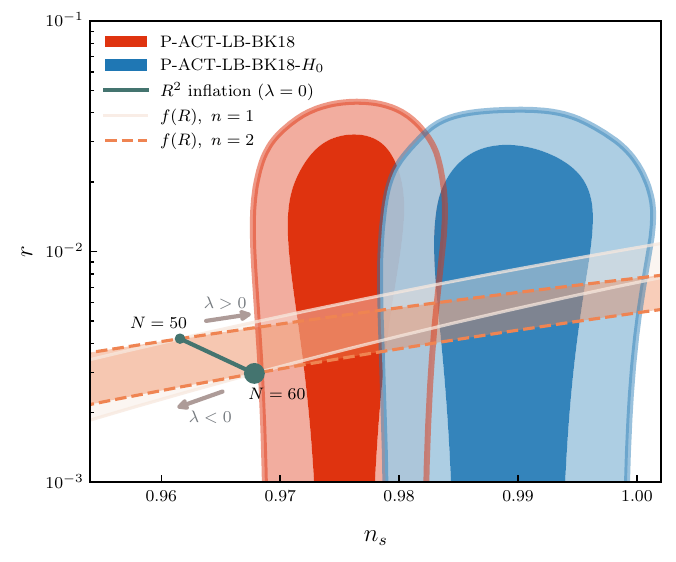}
	\caption{The figure reproduced from Ref.~\citep{Qiu:2025uot} compares the predictions of the non-perturbative exponential $f(R)$ inflation model with constraints obtained from the P-ACT-LB$_{\text{DESI DR2}}$-BK18 combination within the $\Lambda$CDM model and from the corresponding combination supplemented by a SH0ES $H_0$ prior within the EDE framework.}
	\label{fig:fR}
\end{figure}

Another route is to add higher-order curvature corrections to Starobinsky inflation or work within extended $f(R)$-type frameworks~
\citep{Huang:2013hsb,Odintsov:2025jky,Oikonomou:2025htz,Qiu:2025uot}. The non-perturbative exponential $f(R)$ model proposed in Ref.~\citep{Qiu:2025uot}
provides a concrete and minimal realization of this idea: it remains continuously connected to the original Starobinsky model and avoids the corresponding model-dependent tuning while still moving the prediction toward the ACT-preferred larger $n_s$ region. This model is also theoretically motivated. Actually, the Starobinsky inflation has been discussed in connection with swampland criteria~\citep{Lust:2023zql}. On the other hand, in a self-consistent quantum theory of gravity, it is natural to consider curvature corrections beyond the $R^2$ term, and the effects of $R^{3 }$ and $R^{4 }$ corrections on Starobinsky inflation have been investigated in Ref.~\citep{Huang:2013hsb}. The non-perturbative exponential $f(R)$ model is given by
\begin{equation}
	f(R) = R + \frac{R^2}{2\mu^2}\exp\left[-\lambda \left(\frac{R}{\mu^2}\right)^n\right] \,,
\end{equation}
which reduces to the original Starobinsky model in the limit $\lambda \to 0$. The corresponding Einstein frame effective potential is shown in Fig~\ref{fig:fR_potential}. As shown in Fig.~\ref{fig:fR}, a positive $\lambda$ tends to increase $n_s$ relative to the Starobinsky limit and can move the predictions toward larger $n_s$, whereas a negative $\lambda$ tends to lower $n_s$. This model therefore provides a concrete example of how higher-curvature corrections can move plateau-like predictions toward the larger $n_s$ region preferred by ACT DR6 and related CMB/BAO data combinations. 

In addition, analyses of early dark energy (EDE) models using ACT DR6 and DESI DR2 show that EDE can raise the inferred value of $H_0 r_s$ and thereby improve the consistency between CMB and DESI DR2 data~\citep{Poulin:2025nfb}. When a SH0ES prior on $H_0$ is included, the corresponding allowed region is further shifted toward even larger $n_s$ close to the Harrison--Zel'dovich limit $n_s\simeq 1$~\citep{Poulin:2025nfb,Toda:2025kcq,Peng:2025tqt,Pang:2025lvh}. Importantly, the non-perturbative exponential $f(R)$ model remains compatible with this larger $n_s$ region, thereby providing an economical and well-motivated framework for shifting inflationary predictions toward larger $n_s$, as shown in Fig.~\ref{fig:fR}.

Several other model-building directions have also been explored in response to the ACT-motivated shift in $n_s$. These include Higgs/modular and radiatively or quantum-corrected quartic/Higgs-like scenarios~\citep{Aoki:2025wld,Gialamas:2025kef,Yuennan:2025kde}, curvaton and polynomial-potential constructions~\citep{Byrnes:2025kit,Peng:2025bws}, and corrected attractor or quantum-gravity/string-inspired deformations~\citep{Heidarian:2025drk,Wolf:2025ecy,Oikonomou:2025xms}. A detailed comparison of these constructions is beyond the scope of this brief review.

\subsection{Future Observational Probes}

The preceding discussion has focused mainly on $n_s$ and $r$, as inferred from CMB data and CMB$+$BAO combinations. In the coming years, observational tests of inflation are expected to become increasingly multi-probe. For forthcoming CMB observations, improved measurements of CMB $B$-mode polarization from planned satellite missions such as LiteBIRD~\citep{LiteBIRD:2023zmo} and ground-based programs such as the Simons Observatory~\citep{SimonsObservatory:2018koc} and Ali-CPT~\citep{Li:2017drr} are expected to tighten constraints on $r$. At the same time, high-resolution measurements of CMB temperature and polarization anisotropies at high multipoles will further improve constraints on $n_s$ and on possible running of the scalar spectral index, see, e.g., forecast studies for next-generation CMB surveys~\citep{SimonsObservatory:2018koc,CMB-S4:2016ple}.

Ongoing and future LSS observations will also provide complementary probes of inflation. DESI and its extended observations~\citep{DESI:2016fyo}, the Square Kilometre Array (SKA)~\citep{Maartens:2015mra}, and neutral-hydrogen 21-cm surveys are expected to improve, or at least provide complementary constraints on primordial non-Gaussianity and the running of the scalar spectral index~\citep{Meerburg:2019qqi,Camera:2014bwa,Sekiguchi:2017cdy}. Large-volume galaxy surveys can test local-type primordial non-Gaussianity through the scale-dependent bias of tracers~\citep{Dalal:2007cu,Slosar:2008hx,Camera:2014bwa}, while future 21-cm observations can extend the accessible range of wavenumbers and thereby improve sensitivity to possible running of the scalar spectral index~\citep{Sekiguchi:2017cdy,Kohri:2013mxa}. These measurements will therefore provide consistency tests of the single-field slow-roll paradigm that are complementary to those from the CMB alone~\citep{Meerburg:2019qqi}.

\section{Summary and discussion}
\label{Sec_V}

In this paper, we have summarized the basic motivation for inflation, the minimal slow-roll framework, some typical inflationary models, recent observational developments, and the theoretical responses motivated by them. In its simplest realization, canonical single-field slow-roll inflation provides a predictive benchmark, yielding a nearly scale-invariant power spectrum of predominantly adiabatic and approximately Gaussian primordial perturbations~\citep{Mukhanov:1981xt,Hawking:1982cz,Starobinsky:1982ee,Guth:1982ec,Bardeen:1983qw,Maldacena:2002vr}. The typical models discussed in this paper illustrate how the shape and origin of the inflationary potential are reflected in observables such as $n_s$ and $r$. Simple large-field monomial models are strongly constrained by current upper bounds on $r$, whereas small-field models typically predict suppressed tensor amplitudes but are more sensitive to the detailed form of the potential and to the initial conditions near the flat region. Plateau models such as Starobinsky $R^2$ inflation and Higgs inflation instead provide well-motivated examples with robust large-$N_*$ predictions that are consistent with \textit{Planck}-era constraints~\citep{Planck:2018jri}.

Recent ACT DR6 analyses and related CMB$+$BAO data combinations have been interpreted as favoring a larger scalar spectral index $n_{s}$ than that inferred from \textit{Planck} alone~\citep{AtacamaCosmologyTelescope:2025blo,AtacamaCosmologyTelescope:2025nti,Kallosh:2025ijd}, thereby placing conventional plateau models under nontrivial pressure. Although several recent analyses have emphasized that this interpretation should be regarded with caution~\citep{Ferreira:2025lrd,McDonough:2025lzo,Balkenhol:2025wms}, this situation has stimulated a broad range of theoretical responses. In this brief review, we have discussed several mechanisms that can shift inflationary predictions toward larger $n_s$, including a stiffer effective equation of state during reheating, non-minimal couplings to gravity in chaotic inflation~\citep{Kallosh:2025rni}, and the non-perturbative exponential $f(R)$ model~\citep{Qiu:2025uot}. Among them, the non-perturbative exponential $f(R)$ model provides a concise and well-motivated realization of higher-curvature corrections beyond the original Starobinsky $R^2$ framework.

Looking ahead, forthcoming CMB and LSS observations are expected to tighten the constraints on inflationary observables, including $n_s$, $r$, primordial non-Gaussianity, and possible running of the scalar spectral index. Such probes will help determine whether the minimal single-field slow-roll framework remains sufficient or whether extensions are required, and may thereby offer clues about the microphysical origin of inflation.

\begin{acknowledgements}
	This work is supported by the grants from NSFC (Grant No.~12475065, 12547110, 12447101) and the China Manned Space Program with grant no. CMS-CSST-2025-A01.  
\end{acknowledgements}

\bibliographystyle{raa}
\bibliography{refs}

\label{lastpage}

\end{document}